\documentclass[preprint,showpacs,amsmath,amssymb]{revtex4}

\usepackage{graphicx}
\usepackage{dcolumn}
\usepackage{bm}
\newcommand{\etal}{\emph{et al.}}
\newcommand{\be}{\begin{equation}}
\newcommand{\ee}{\end{equation}}
\newcommand{\bfig}{\begin{figure}}
\newcommand{\efig}{\end{figure}}
\newcommand{\incl}{\includegraphics}

\begin{document}      

\title{Unusual Hall Effect Anomaly in MnSi Under Pressure}

\author{Minhyea Lee$^1$$^{\dagger}$, W.  Kang$^2$, Y. Onose$^3$ Y. Tokura$^{3,4}$ and N. P. Ong$^1$
}
\affiliation{
$^1$Department of Physics, Princeton University, Princeton, NJ 08544, USA\\
$^2$Department of Physics, University of Chicago, Chicago, IL 60637, USA\\
$^3$Department of Applied Physics, University of Tokyo, Tokyo 113-8656, Japan
$^4$ERATO, JST, Spin Superstruture Project (SSS), Tsukuba 305-8562, Japan
}
\date{\today}      
\begin{abstract}
We report the observation of a highly unusual Hall current 
in the helical magnet MnSi in an applied pressure $P$ = 6--12 kbars.  
The Hall conductivity displays a distinctive step-wise field profile 
quite unlike any other Hall response observed in solids.
We identify the origin of this Hall current with 
the effective real-space magnetic field due
to chiral spin textures, which may be a precursor of the partial-order state at $P>$14.6 kbar.  We discuss evidence favouring the chiral spin
mechanism for the origin of the observed Hall anomaly.
\end{abstract}

\maketitle                   
In the helical, itinerant magnet MnSi, 
a novel magnetic state with ``partial order'' 
has been reported by
Pfleiderer and coworkers  
above a critical applied pressure $P_c$= 14.6 kbar~\cite{Pfleiderer01,Pfleiderer04,Pfleiderer07}.  
The neutron diffraction intensity 
is broadly distributed over the surface of a sphere in momentum space,
but is resolution limited in the radial direction.  
Several groups have proposed that the state harbors
non-trivial topological spin textures~\cite{Tewari,Binz1,Rossler,Fischer}.
We have observed a highly unusual Hall current in MnSi
at pressures (6--12 kbar) just below $P_c$, which 
appears to be caused by fluctuations into the 
chiral spin state at temperatures $T$ near 
the Curie transition temperature $T_C$.

At ambient pressure, MnSi, which is non-centrosymmetric with
the crystal structure B20, undergoes a transition at $T_C\simeq$ 30 K to a helical 
state with a long pitch $\lambda\sim$ 180 \AA. The 
wavevector ${\bf q}$ is weakly pinned along the
$\langle 111\rangle$ direction.
The helical state reflects the competition between the
exchange energy and the Dzyaloshinky-Moriya term
~\cite{Ishikawa84,Moriya,Pfleiderer97}.
In a magnetic field $\bf H$, $\bf q$
shifts to alignment, and the helical state evolves to a conical
magnetic state, whose cone angle steadily decreases to zero at 
a field $H_s\sim$0.6 T.
Under pressure, $T_C$ decreases monotonically, reaching zero 
at the critical pressure $P_c\sim$14.6 kbar (Fig. \ref{figphase}, inset).  Above $P_c$, the ``partial-order'' state displays
a non-Fermi liquid exponent in its resistivity ~\cite{Pfleiderer01} in addition to
the unusual neutron diffraction spectrum.

The new Hall anomaly is observed in weak $H$ in the pressure 
interval 6 $<P<$ 12 kbar below the 
curve of $T_c$ vs. $P$ (shaded region in the inset of 
Fig. \ref{figphase}). Figure \ref{figphase} (main panel) displays the Hall resistivity $\rho_{yx}$ measured at several
temperatures ($T$) with $P$ fixed at 11.4 kbar ($T_C$ = 11.3 K).  
At the lowest $T$ (0.35 and 2.5 K), $\rho_{yx}$ is 
hole-like and $H$-linear.  Between 5 and 10 K, however,
we observe a prominent Hall anomaly with a most unusual profile.
The step-like onset at the field $H_1\sim$0.1 T and the equally
abrupt disappearance at $H_2\sim$0.45 T stands in sharp contrast
with the broad background at higher $H$ [the latter is the {\it conventional} anomalous Hall effect (AHE) term common to all ferromagnets].  
We show below that this unusual anomaly is the Hall-current response
produced by coupling between the spin of charge carriers 
with chiral spin textures.  Its observation
provides strong evidence that the chiral spin textures exist
over a large fraction of the phase diagram for $P<P_c$.

Motivated by the results of Refs. \cite{Pfleiderer01,Pfleiderer04,Pfleiderer07},
several groups~\cite{Binz1,Rossler,Tewari,Fischer}
have proposed states comprised of crystalline arrays of 
magnetic textures.  In Ref.~\cite{Binz1}, the transition at $P_c$ 
is from a single-spiral state to a bcc ``spin crystal'' 
comprised of 6 spirals.  In Refs.~\cite{Rossler,Fischer},
the proposed state is either a square lattice
configuration of topological defects called skyrmions~\cite{Rossler,Fischer}, or 
a cubic network of line defects which are double-twist 
configurations~\cite{Fischer}.
The textures are closely related to the defects
previously studied in the blue phase of nematic liquid crystals~\cite{Tewari,Fischer}.

Hall measurements were made in a $^3$He cryostat using a
miniature clamp-type pressure cell (13 mm dia.) made of BeCu alloy and tungsten carbide with Fluorinert (FC-77) as the pressure medium.  At low $T$, the pressure was calibrated by the superconducting transition of a Pb coil detected by AC
susceptibility.  
The high-purity MnSi crystals (of resistivity ratio $\rho$(300K)/$\rho$(4.2K) $\simeq$ 60 and size 1.1$\times$0.5$\times$0.060 mm$^3$) were cut from 
boules grown in a floating-zone furnace.   
The Hall voltage $V_H$ was checked to scale linearly with current $I$ (typically
0.5 to 1 mA).  At each $T$ and $P$, $V_H$ was recorded with $H$ swept 
at the rate 0.02 T/min. in the sequence $0\rightarrow -1 T\rightarrow 0\rightarrow 1 T\rightarrow 0$ to eliminate errors from induced emf's and drifts 
in $T$. The observed Hall anomaly is always antisymmetric in $H$ even without antisymmetrization of the 4 raw curves.  
Results obtained in the 2 crystals investigated are closely similar.

In ferromagnets, the observed Hall resistivity $\rho_{yx}$ is the sum of
the ordinary Hall resistivity $\rho_{yx}^N =  \sigma_{xy}^N\rho^2$ and the AHE term $\rho_{yx}^A = \sigma_{xy}^A\rho^2$, where 
$\rho = \rho_{xx}$ is the resistivity and $\sigma_{xy}^N$
and $\sigma_{xy}^A$ are the ordinary and anomalous 
Hall conductivities, respectively.  Dividing by $\rho^2$, we have
\be
\frac{\rho_{yx}}{\rho^2} = \sigma_{xy}^N + \sigma_{xy}^A.
\label{rhoxy}
\ee
The first term is strictly $H$ linear in low $H$, while the second term 
scales as the uniform magnetization $M(T,H)$.

To bring out the surprising nature of the new anomaly, 
we recall the salient features of the Hall Effect 
at ambient $P$~\cite{Aeppli,Lee}.  
In high-purity MnSi, 
Lee \etal~\cite{Lee} have shown that, in spite of
the large magnetoresistance (MR), the
$H$ dependence of $\sigma_{xy}^A$ at ambient $P$ strictly mimics 
that of $M(T,H)$, viz. $\sigma_{xy}^A(T,H) = S_H M (T,H)$,
with $S_H$ a constant independent of $T$ and $H$.  
This scaling confirms
a key prediction of the Karplus-Luttinger (KL) theory~\cite{Karplus}
(and its generalization using the Berry phase~\cite{Niu,Nagaosa,Jungwirth}).

\bfig[h]			
\incl[width=7cm]{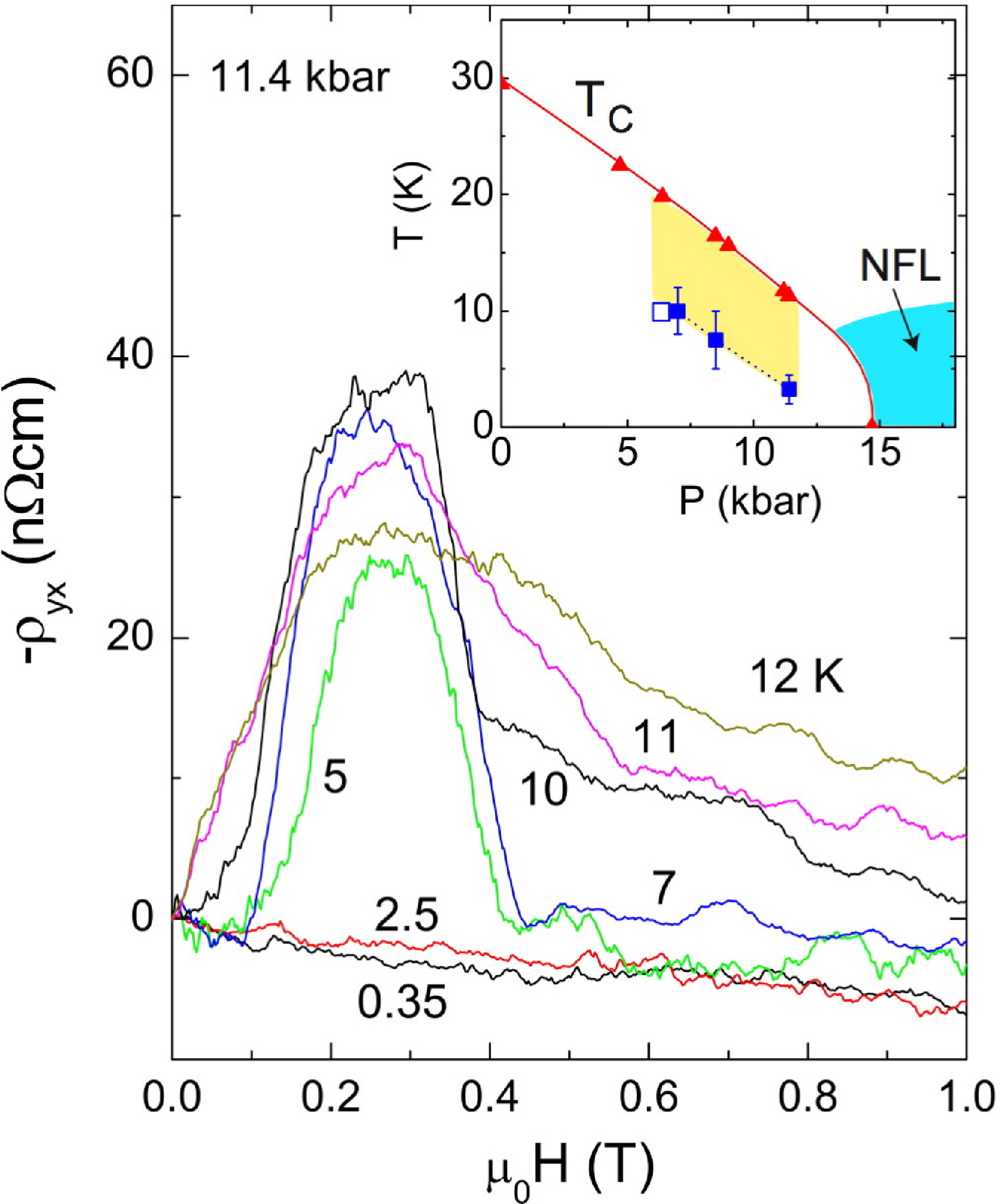}
\caption{\label{figphase} (color online)
(Main panel) Curves of -$\rho_{yx}$ in MnSi at 11.4 kbar revealing a large
Hall anomaly in the field range 0.1$< H <$0.45 T for several $T<T_C$ ($T_C=$11.3 K),
with $\bf H$ nominally along (111).  The anomaly (electron-like in sign) arises from a new contribution $\sigma_{xy}^C$ 
to the total Hall conductivity.  In the phase diagram (inset), the
shaded region is where $\sigma_{xy}^C$ is resolved. $T_C$ was determined
from $\rho$ vs. $T$.  Data from Samples 1 and 2 are shown as solid and
open squares, respectively.  The non-Fermi liquid (NFL) region is shaded blue
(adapted from Ref.~\cite{Pfleiderer04}).
}
\efig


\bfig[h]			
\incl[width=8.5cm]{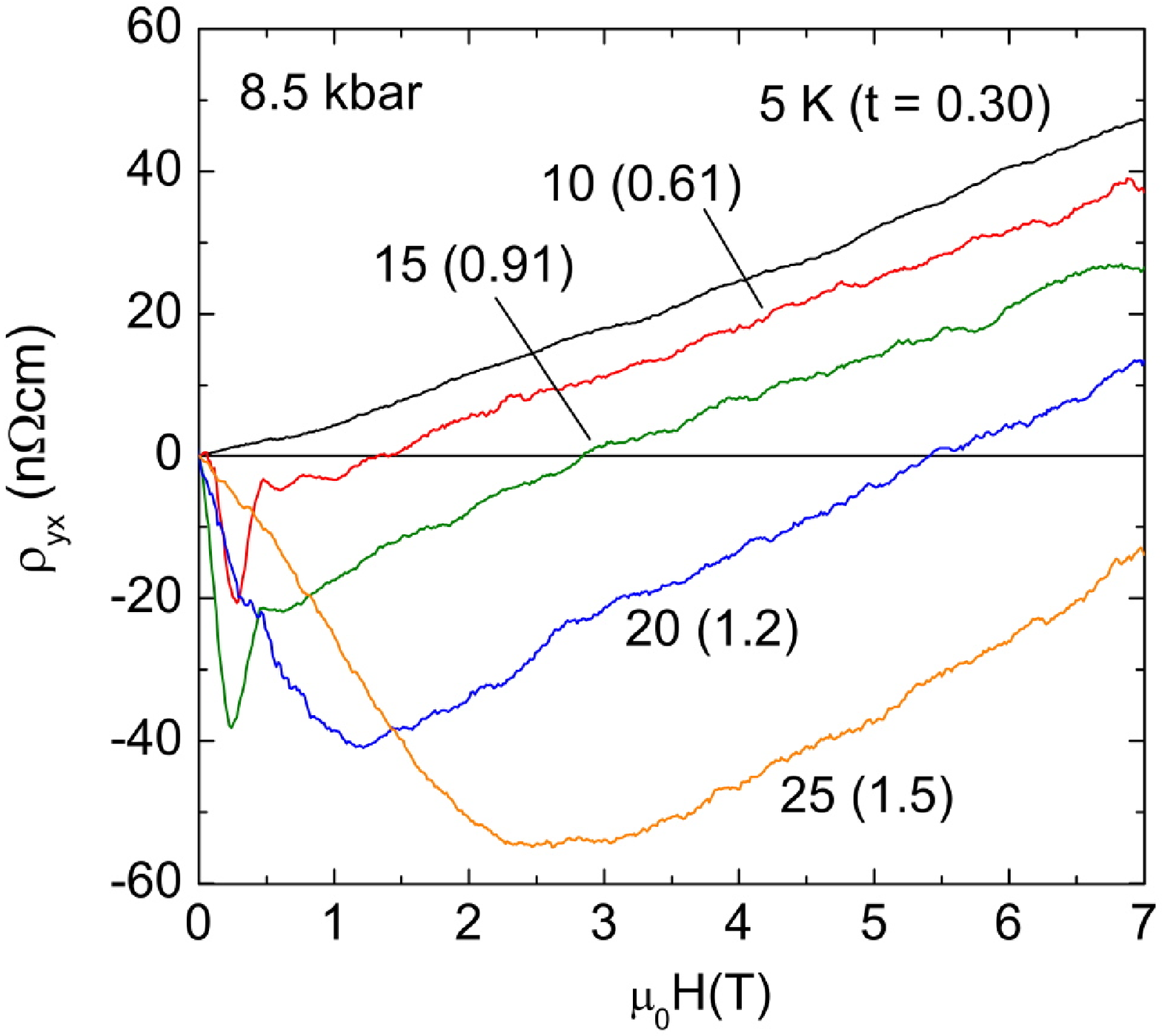}
\caption{\label{fighigh} (color online)
Curves of $\rho_{yx}$ extending to 7 T showing the relation of the 
Hall anomaly to other Hall terms 
(at the applied $P$ = 8.5 kbar, $T_C$ = 16.4 K).
At 5 K, the hole-like $\sigma_{xy}^N$ dominates $\rho_{yx}$
to produce an $H$-linear background. At $T$ = 10 and 15 K, 
at which the Hall anomaly is observed as sharp negative spikes, the 
AHC term $\rho_{yx}^A= \sigma_{xy}^A\rho^2$ grows in prominence
as $\rho$ increases.  Above $T_C$ 
(20 and 25 K), the Hall anomaly vanishes, while $\rho_{yx}^A$
continues to increase in magnitude.
The reduced temperature $t=T/T_C$ is 
shown in parenthesis for each curve.
}
\efig

Returning to the Hall curves in MnSi under pressure, 
we plot in Fig. \ref{fighigh} curves of $\rho_{yx}$ 
over a broader field range, with $P$ fixed at 8.5 kbar 
(at which $T_C$ = 16.4 K).  
Starting at 5 K, we observe that $\rho_{yx}$ is linear in $H$ to 7 T.
This reflects the dominance of $\sigma_{xy}^N\sim\ell^2$ 
in the limit of large mean-free-path $\ell$.  With increasing $T$, 
however, the rapid increase of $\rho$ strongly amplifies the anomalous 
term $\rho_{yx}^A$ which emerges as a negative contribution with a 
broad shoulder (e.g. at 2.5 T at 25 K).  We refer to this
term as the ``conventional'' AHE term.  At 10 and 15 K, we see the 
emergence of the new Hall anomaly as a sharp negative spike in weak fields.  Raising $T$ above $T_C$ 
removes the spike (curves at 20 and 25 K).  The curves 
above $T_C$ are closely similar to those
observed at ambient $P$ (where the spike is absent).

The new Hall anomaly is qualitatively distinct from the Lorentz-force term
and the KL term.  As mentioned, theory predicts
that coupling of the carrier spin to local textures 
of $\bf M(r)$ produces a large anomalous Hall current via the Berry phase~\cite{Ye,Tatara}.
In the past decade, the Berry-phase approach has greatly
extended the purview of the KL theory~\cite{Niu,Nagaosa,Jungwirth}.  
In a periodic lattice, the ``overlap'' of wave 
functions $u_{\bf k}$ defines the Berry gauge potential 
${\bf A(k)} = \langle u_{\bf k}|i\nabla_{\bf k}|u_{\bf k}\rangle$,
whose curl gives an effective magnetic field $\bf \Omega(k)$ that lives
in $\bf k$ space.  When $\bf \Omega(k)$ is rendered finite (by  breaking time-reversal invariance), it leads to orbit deflection in $\bf k$ space, to reproduce the KL term $\sigma_{xy}^A$ in Eq. \ref{rhoxy}.

Quite distinct from this strictly orbital coupling, the Berry phase can 
produce an \emph{additional} effective magnetic field ${\bf B}_{\phi}$ 
via the spin degrees of freedom.  
The spin-mediated mechanism was initially invoked to explain 
AHE experiments in manganites~\cite{Matl} and pyrochlores~\cite{Taguchi}.
We consider the simplest example in which 
the spin $\bf s$ of a hopping electron aligns
by Hund coupling 
$J_H$ to the ion's local moment ${\bf S}_i$ 
at each site $i$~\cite{Matl,Taguchi}.  
As the electron completes a closed
loop linking 3 non-coplanar spins ${\bf S}_i$ ($i = 1,2,3$), 
$\bf s$ describes a cone of finite solid angle $\Omega_s$ (Fig. \ref{figsxy}, inset).
Hence, the electron acquires  a Berry phase $\phi_B = \frac12\Omega_s$,
which translates to a magnetic
field in real space $B_\phi = (\phi_B/2\pi)(\phi_0/{\cal A})$ 
that can be extremely large (${\cal A}$ is the loop area and 
$\phi_0$ the flux quantum). For e.g., 
even for $\Omega_s\sim\pi/100$ over an
area ${\cal A}\sim 5\times 5$ \AA$^2$, we have $B_\phi\sim$42 T.
In turn, ${\bf B}_{\phi}$ produces a large Hall 
conductivity $\sigma_H^C$ that is proportional to the 
chirality $\chi_c = {\bf S}_1\cdot{\bf S}_2\times{\bf S}_3$~\cite{Taguchi,Ye,Tatara}.

The foregoing also applies to itinerant ferromagnets~\cite{Tatara}.
In a spiral helimagnet such as MnSi, 
the local spin direction $\bf S(r)$ varies periodically
with a pitch set by the wavevector $\bf q$.
For an itinerant electron,
the exchange energy forces its spin $\bf s$ to follow
the spatial variation of $\bf S(r)$.  
However, as emphasized in Ref.~\cite{Binz2},
a spiral state with a single $\bf q$ 
has zero chirality.  In order 
to produce finite chirality, we must have a 
multi-$\bf q$ spiral state, as has been 
proposed for the partial order state for $P>P_c$.

In the novel states proposed~\cite{Binz1,Rossler,Tewari,Fischer}, 
the presence of skyrmions or double-twist configurations
naturally leads to chirality.  
We expect $\sigma_H^C$ to be proportional to the skyrmion 
number $N_s = \int d^3r \Phi_z({\bf r})$, with the skyrmion density 
$\Phi_z = (8\pi)^{-1} {\bf\hat{n}}\cdot (\partial_{x}{\bf\hat{n}}\times\partial_{y}{\bf\hat{n}})$ and
$\bf\hat{n} = S(r)/|S(r)|$.

For the region of our Hall experiment ($P<P_c$), 
we expect fluctuations towards the 
multi-$\bf q$ state to be favorable at temperatures just 
below $T_C$ (the fluctuations are strongly suppressed
as $T\rightarrow 0$).  Hence, within the broad swath
in which $\sigma_{xy}^C$ is observed
(Fig. \ref{figphase} inset), we propose that 
chirality exists caused by strong
fluctuations into multi-$\bf q$ helical states.

There is considerable evidence for strong fluctuations
in the region $P<P_c$.
The muon spin rotation data by Uemura \etal~\cite{Uemura} show
that, for $P<P_c$, magnetic order exists only in a partial volume fraction, 
in agreement with conclusions from nuclear magnetic resonance~\cite {Yu} and neutron scattering~\cite{Fak05} experiments.  In a broad range of $P$
below $P_c$, strong fluctuations appear as a precursor to the
partial-ordered state above $P_c$.

\bfig[h]			
\incl[width=8.5cm]{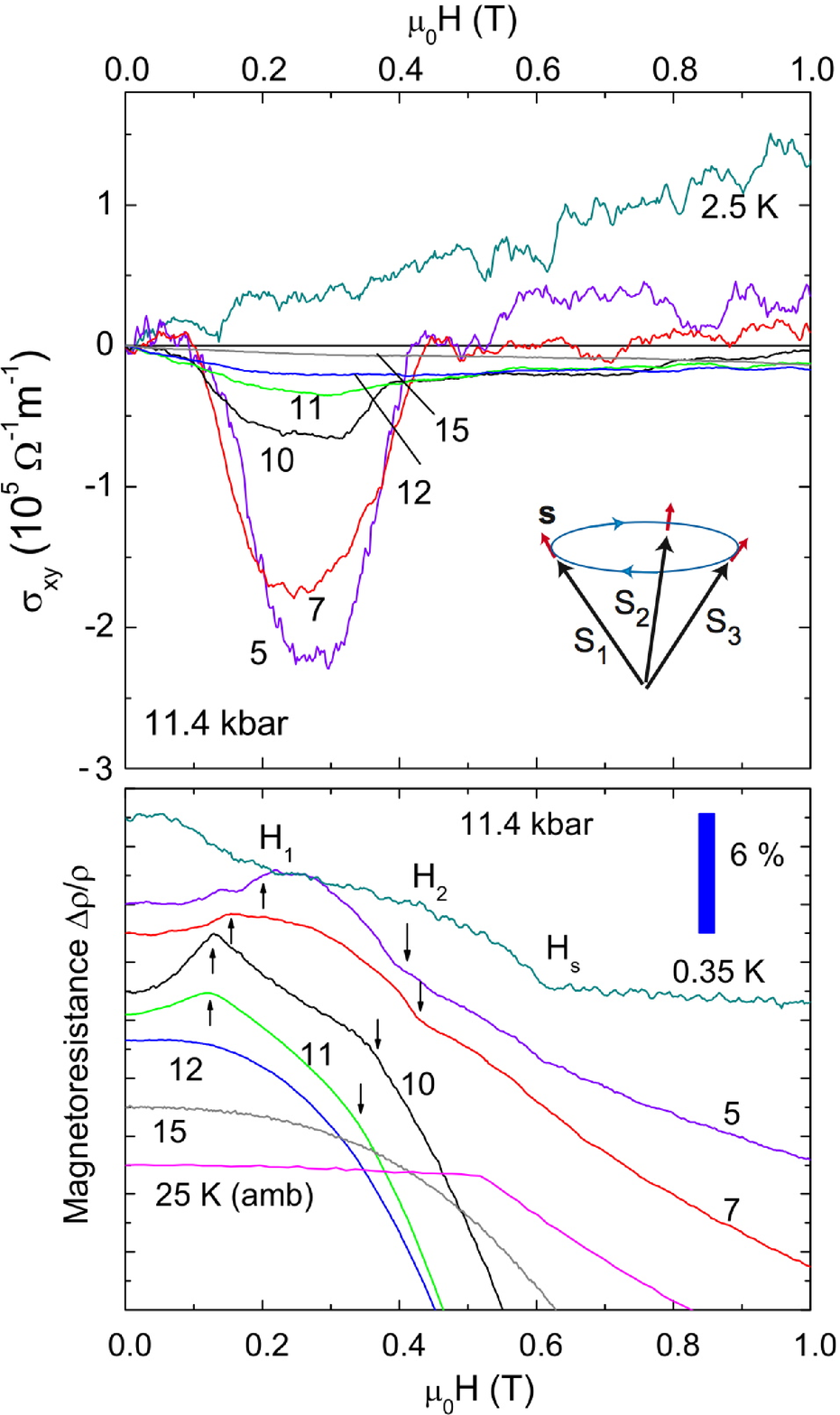}
\caption{\label{figsxy} (color online)
Panel (a) 
Curves of $\sigma_{xy}$ vs. $H$ at $P$ = 11.4 kbar
at selected $T$.  The Hall anomaly $\sigma_{xy}^C$ is
prominent between 11 and 5 K between the fields $H_1$ 
and $H_2$.  At 2.5 K, only the normal term $\sigma_{xy}^N$ is
observed. The inset shows the itinerant spin $\bf s$ aligned
with the local moment ${\bf S}_i$ on site $i$ by a large $J_H$.  
As the electron closes the path 1-2-3-1, 
it acquires a Berry phase $\phi_B$.
Panel (b)
Curves of the MR $\Delta\rho/\rho$ vs. $H$ at selected $T$
at 11.4 kbar (curves are offset for clarity).  Below $T_C$ (11.3 K)
anomalies indicate fields $H_1$ (uparrows), $H_2$ (downarrows)
and $H_s$.  By comparison, the MR is nearly zero below $H_s$ = 0.6 T at 
ambient pressure (thin curve at 25 K).
}
\efig

Within this picture, we may understand several puzzling features
of the Hall data (as well as the MR).
To examine these issues in more detail, we transform $\rho_{yx}$ to 
$\sigma_{xy} = \rho_{yx}/\rho^2$ using the simultaneously 
measured curve of $\rho$ vs. $H$.  
As plotted in Fig. \ref{figsxy}a, the Hall anomaly is now apparent
as a large Hall conductivity $\sigma_{xy}^C$, with a distinctive
field profile.   
Just below $T_C$, the anomaly first appears as a shallow 
hull feature (compare curves at 12 and 11 K). 
With decreasing $T$, it deepens considerably.  
Comparing the curves of $\sigma_{xy}$
with the MR curves, we see that $\sigma_{xy}^C$ is restricted to the 
narrow interval $H_1<H<H_2$.
At 5 K, its magnitude is $\sim$10 times larger
than the other 2 Hall terms, and $\sim 1\%$ of the zero-$H$ 
conductivity $\sigma$.  Finally, at 2.5 K, the anomaly
vanishes, leaving an $H$-linear background that is
dominated by $\sigma_{xy}^N$.

We next discuss the evidence for the chiral-spin mechanism.
First, we note that, between $H_1$ and $H_2$, 
$|\sigma_{xy}^C|$ attains remarkably large values.  
At  ambient pressure 5 K with $H$ = 0.3 T, the measured values of the
ordinary term and the KL term, $\sigma_{xy}^N$ and $\sigma_{xy}^A$, are
$\sim +3.2\times 10^4$ and $-1.2\times 10^4$ ($\Omega_s$m)$^{-1}$, 
respectively~\cite{Lee}.  By comparison, 
$|\sigma_{xy}^C|$ is 10 times larger
than either of these values.  Such a large $\sigma_{xy}^N$ 
is difficult to understand with orbital mechanisms given
the small values of $H_1$ and $H_2$.  By contrast, 
the chiral spin 
term is easily capable of producing such a large Hall response.
The effective field $B_{\phi}$ can exceed 40 T despite
the small applied $H$.

Secondly, $\sigma_{xy}^C$ is seen to be finite only
within a very narrow field interval $(H_1,H_2)$.  The
relatively abrupt vanishing of $\sigma_{xy}^C$ 
at $H_2\sim$0.45 T is striking.  Such an abrupt vanishing
of the Hall current seems impossible to 
realize with orbital 
mechanisms (carrier mobilities cannot be abruptly changed
in such weak $H$).  By contrast, the spin-Berry phase model
applied to MnSi anticipates that $\sigma_{xy}^C$
must vanish at a field below $H_s$, as the 
cone angle of the spiral state is suppressed to
zero.  In the ferromagnetic state
above $H_s$, spin textures are energetically
prohibitive.  As the spin textures are removed with
increasing $H$, $\sigma_{xy}^C$ must vanish.  Hence $H_s$ 
represents an upper bound for a finite $\sigma_{xy}^C$.
The Hall results show that the vanishing actually
occurs at the slightly lower field $H_2$.  
The observed
mesa profile of $\sigma_{xy}^C (H)$ 
has been reproduced in a recent calculation~\cite{Binz2} based
on the bcc1 ``spin-crystal'' ground state of MnSi.

Thirdly, the picture described also clarifies the origin of
the kink anomalies long known in MnSi
under pressure~\cite{Kadowaki}.  The low-$T$ transverse MR of 
MnSi is very large and negative throughout its phase diagram
because of its long $\ell$.  Under pressure, weak
anomalies appear at fields $H<H_s$.  
Figure \ref{figsxy}b compares the transverse MR measured in our sample 
at ambient pressure and at 11.4 kbar 
with $\bf H\perp I$ (applied current).  First, we examine
the MR curve at ambient $P$ (lowest curve,
at 25 K, with $T_C$ = 30 K).  Surprisingly, the MR is almost zero 
below $H_s$.  A sharp kink at $H_s$ signals saturation of the moments,
followed by a steep decrease of $\rho$ at larger $H$.  The absence of
MR implies that, as $\bf q$ re-orients in $\bf H$ and the cone angle decreases
~\cite{Ishikawa},
there is no change in the carrier scattering rate $\Gamma$ in $H<H_s$ ~\cite{Lee, Kadowaki}.  Consequently, the 
changes below $H_s$ involve no change in magnetic disorder or the 
creation of spin defects at ambient $P$.  
The ``rigidity'' of the spiral state at ambient 
pressure also explains why the Hall anomaly is not observed at
ambient pressure.

By contrast, at $P$ = 11.4 kbar, the MR exhibits kinks in 
the interval $0<H<H_s$.  Above $T_C$ (11.3 K), the MR
decreases smoothly.  At 11 K, the 2 field scales $H_1$ 
and $H_2$ inferred from $\sigma_{xy}$ (Fig. \ref{figsxy}b) 
become apparent (up and down arrows, respectively).  Their
positions change only slightly with $T$. However, they
become more sharply defined as $T$ approaches $T_C$ from below.
Throughout the interval (0,$H_s$), the visible MR anomalies 
imply that changes in the magnetic structure are accompanied by the 
production of magnetic defects and textures which increase $\Gamma$.
Hence, in both transport channels, the onset and disappearance
of $\sigma_{xy}^C$ at 11.4 kbar at $H_1$ and $H_2$ are
nearly coincident with the MR anomalies.

Our experiment reveals that, when a charge current flows 
through a region with chiral spin textures, a large Hall current 
appears.  The spin-texture generated Hall current disappears 
if the textures are erased by increasing $H$. Because of the abruptness of 
its onset and disappearance in a narrow field interval, the 
new Hall conductivity $\sigma_{xy}^C$ is easily distinguished
from both the conventional AHE term and Lorentz-force term.  
Its distinctive profile suggests that 
it may serve as a sensitive detector of chiral spin textures 
in helical magnets.  We find
that, in MnSi, these textures exist over a significant region of the
phase diagram below the $T_c$ curve for $P<P_c$.

We thank B. Binz,  A. Vishwanath and M. Hermele for 
valuable discussions.
The research at Princeton is 
supported by the U.S. National Science Foundation
under MRSEC Grant DMR 0213706.

$^{\dagger}$\emph{Present address of ML}: National Institute of Standards and Technology, Boulder, CO 80305.


\begin{thebibliography}{}

\bibitem{Pfleiderer01}  C. Pfleiderer,  S. R. Julian and G. G. Lonzarich,   
{ Nature}  {\bf 414}, 427- 429 (2001).

\bibitem{Pfleiderer04}  C. Pfleiderer \etal,  
{ Nature}  {\bf 427}, 227- 231 (2004).

\bibitem{Pfleiderer07}  C. Pfleiderer, P. B\"{o}ni, T Keller, U.K. R\"ossler and A. Rosch,    
{ Science} {\bf 316}, 1871 (2007).


\bibitem{Rossler}  U. K. R\"ossler, U. N. Bogdanov and  C. Pfleiderer, 
{Nature}  {\bf 442}, 797- 801 (2006).


\bibitem{Fischer} 
I. Fischer, N. Shah, and A. Rosch, \prb {\bf 77}, 
024415 (2008).


\bibitem{Tewari}  S. Tewari, D. Belitz and T. R. Kirkpatrick,     
\prl {\bf 96}, 047207 (2006).

\bibitem{Binz1} B. Binz, A. Vishwanath, and V. Aji, \prl {\bf 96}, 207202 (2006); B. Binz and  A. Vishwanath, \prb {\bf 74},214408 (2006). 




\bibitem{Ishikawa84} Y. Ishikawa and M. Arai,  {\it J. Phys. Soc. Jpn} {\bf 53} 2726-2733, (1984).
\bibitem{Moriya}  {\it Spin Fluctuation in Itinerant Electron Magnetism}, T. Moriya, Springer Series in Solid-State Sciences, Springer, Berlin (1985);  and references therein.
\bibitem{Pfleiderer97} C. Pfleiderer, G. J. McMullan, S. R. Julian and G. G. Lonzarich, \prb {\bf 55}, 8330 (1997).

\bibitem{Aeppli}  N. Manyala,  {\it et al.}  
{Nature Materials} {\bf 3}, 255-262 (2004).

\bibitem{Lee} M. Lee, Y. Onose, Y. Tokura and N. P. Ong, 
\prb {\bf 75}, 172403 (2007).


\bibitem{Karplus} R. Karplus and J. M. Luttinger, Phys. Rev. {\bf 95}, 1154 (1954).

\bibitem{Niu} Ganesh Sundaram and Qian Niu, Phys. Rev. B {\bf 59}, 14915 (1999).

\bibitem{Nagaosa} M. Onoda, N. Nagaosa, J. Phys. Soc. Jpn. {\bf 71}, 19 (2002).

\bibitem{Jungwirth} T. Jungwirth, Qian Niu, A. H. MacDonald,
Phys. Rev. Lett. {\bf 88}, 207208 (2002).



\bibitem{Ye} J. Ye \etal, \prl {\bf 83}, 3737 (1999).

\bibitem{Tatara} Gen Tatara and Hikaru Kawamura, Jnl. Phys. Soc. Jpn., 
{\bf 71}, 2613 (2002).

\bibitem{Matl} P. Matl \etal, \prb {\bf 57}, 10248 (1998).

\bibitem{Taguchi} Y. Taguchi, Y. Oohara, H. Yoshizawa, N. Nagaosa,
Y. Tokura, Science {\bf 291}, 2573 (2001).

\bibitem{Binz2} B. Binz and A. Vishwanath, Physica B {\bf 403}, 1336 (2008).

\bibitem{Uemura}  Y. J. Uemura {\it et al.},    
{ Nature Physics} {\bf 3}, 29 (2007).

\bibitem {Yu} W. Yu \etal, 
\prl {\bf 92} 086403 (2005).


\bibitem{Fak05}  B. F\aa k, R. A. Sadykov, J. Flouquet and G. Lapertot,   
{J. Phys.: Cond. Matter} {\bf 17} 1635-1644 (2005).

\bibitem{Kadowaki}K. Kadowaki, K. Okuda and M. Date, {J. Phys. Soc. Jpn.} {\bf 51}, 2433 (1982).

\bibitem{Ishikawa} Y. Ishikawa, G. Shirane, J. A. Tarvin
and M. Kohgi, Phys. Rev. B {\bf 16}, 4956 (1977).


\end{thebibliography}
\end{document}